\documentclass[aps,prd,twocolumn,showpacs,nofootinbib]{revtex4-1}

\usepackage{graphicx}
\usepackage{amsmath}
\usepackage{amssymb}
\usepackage{bm}
\usepackage{bbm}
\usepackage{color}
\usepackage{slashed}
\usepackage[colorlinks=true,linkcolor=blue,citecolor=blue, urlcolor=blue]{hyperref}

\begin{document}

\title{Next-to-leading order QCD corrections to the production of $B_c$ and $B_c^*$ through $W^+$-boson decays}

\author{Xu-Chang Zheng$^{a}$}
\email{zhengxc@cqu.edu.cn}
\author{Chao-Hsi Chang$^{b,c,d}$}
\email{zhangzx@itp.ac.cn}
\author{Xing-Gang Wu$^{a}$}
\email{wuxg@cqu.edu.cn}
\author{Jun Zeng$^{a}$}
\email{zengj@cqu.edu.cn}
\author{Xu-Dong Huang$^{a}$}
\email{hxud@cqu.edu.cn}

\affiliation{$^a$ Department of Physics, Chongqing University, Chongqing 401331, P.R. China.\\
$^b$ Key Laboratory of Theoretical Physics, Institute of Theoretical Physics, Chinese Academy of Sciences, Beijing 100190, China.\\
$^c$ School of Physical Sciences, University of Chinese Academy of Sciences, Beijing 100049, China.\\
$^d$ CCAST (World Laboratory), Beijing 100190, China}

\begin{abstract}

In this paper, we calculate the total decay widths for the $W^+$-boson decays, $W^+ \to B_c+b+\bar{s}+X$ and $W^+ \to B^*_c+b+\bar{s}+X$, up to next-to-leading order (NLO) accuracy within the framework of the nonrelativistic QCD theory. Both the fixed-order and the fragmentation approaches are adopted to do the calculation. Differential decay widths $d\Gamma/dz$ and $d\Gamma/ds_1$ are also given. We find that the NLO corrections are significant in those two $W^+$ decay channels. Our numerical results show that at the LHC, there are about $7.03\times 10^4$ $B_c$-meson events and $5.10\times 10^4$ $B^*_c$-meson events to be produced via the $W^+$-boson decays per operation year.

\end{abstract}


\maketitle

\section{Introduction}
\label{secIntro}

The $(c\bar{b})$-quarkonium is a unique system in the Standard Model (SM) which carries two different heavy flavors. Studies on its production, decay and mass spectrum and etc. provide us a good platform to understand the strong and weak interactions deeply. The ground state $B_c$ meson was first observed by the CDF collaboration at the Tevatron \cite{CDF} and it attracts lots of interests since then. At present, the direct production of $B_c$ meson and its excited states have been studied extensively in $pp$~\cite{ppBc1, ppBc2, ppBc3, ppBc4, ppBc5, ppBc6, ppBc7, ppBc8, ppBc9, ppBc10, ppBc11, ppBc12, ppBc13, ppBc14, ppBc15, ppBc16, ppBc17}, $e^+e^-$~\cite{eeBc1, eeBc2, eeBc3, eeBc4, Bcfragnlo, eeBc5} and $ep$~\cite{epBc1,epBc2} collisions.

Besides the direct production mechanisms, the $B_c$ meson can also be indirectly produced through the top-quark~\cite{topBc1, topBc2, topBc3}, the $Z^0$-boson~\cite{fragbc1, Z0Bc1, Z0Bc2, Z0Bc3, Z0Bc4}, the Higgs-boson~\cite{HBc1, HBc2} and the $W$-boson~\cite{WBc1, WBc2, WBc3} decays. These indirect production channels can also generate abundant $B_c$ mesons at the LHC or the future high-energy colliders. The $W$-boson is the propagating media for the weak interaction, and the study on it is important for testing the SM. The LHC is a fruitful $W$-boson factory, there are about $3.07 \times 10^{10}$ $W$-bosons to be produced at the LHC per operation year~\cite{WBc1}. In the paper, we shall concentrate on the production of the $B_c$ meson and its first excited state $B^*_c$ meson through the $W$-boson decays.

The heavy constituent quarks move non-relativistically in the $B_c$ meson, and the processes involving the $B_c$ meson can be calculated within the nonrelativistic QCD (NRQCD) factorization formalism~\cite{nrqcd}. Generally, the production cross-section or the decay width can be factorized into the product of the short-distance coefficients and the long-distance matrix elements. The short-distance coefficients describe the production or decay rate of the heavy quark-antiquark pair, which can be perturbatively calculated in powers of the strong coupling constant $\alpha_s(m_Q)$. The non-perturbative long-distance matrix elements describe the formation of the $B_c$ meson from the heavy quark-antiquark pair, which can be calculated through potential models or lattice QCD.

The excited states of the $B_c$ meson shall directly or indirectly decay to the ground state $B_c$ meson via electromagnetic or strong interactions with $\sim100\%$ probability, so these excited states are important sources of the $B_c$ meson. Moreover, the production of the excited state is also interesting by itself. So, in addition to the $B_c$ meson production, we shall also consider the production of the spin-triplet $^3S_1$ state $B_c^*$. The production of $B_c$ and $B^*_c$ mesons via the $W$-boson decays at the leading order (LO) level has been studied in Refs.\cite{WBc1, WBc2}. Since the masses of $b$ and $c$ quarks are not too large compared to the QCD asymptotic scale $\Lambda_{\rm QCD}$, the higher-order QCD corrections could be important. In this paper, we shall study the $B^{(*)}_c$ meson production via the $W$-boson decays up to next-to-leading order (NLO) accuracy.

There are two decay channels for the inclusive $B_c$ production through the $W^+$-boson decays at ${\cal O}(\alpha_s^2)$, i.e., $W^+ \to B_c+b+\bar{s}$ and $W^+ \to B_c+c+\bar{c}$. However, due to the small value of $\vert V_{cb} \vert$, the contribution from the decay channel $W^+ \to B_c+c+\bar{c}$ is depressed. More explicitly, $\Gamma_{W^+ \to  B_c +c+\bar{c}}/\Gamma_{W^+ \to  B_c +c+\bar{s}}=0.086$ and $\Gamma_{W^+ \to  B_c^*+c+\bar{c}}/\Gamma_{W^+ \to  B_c^* +c+\bar{s}}=0.150$ at the LO level can be derived from the results of Ref.\cite{WBc2}. Thus, in this paper, we only consider the contributions from the decay channels $W^+ \to B^{(*)}_c+b+\bar{s}+X$.

Any physical observable is independent of the renormalization scale, but there is renormalization scale ambiguity for the fixed-order pQCD predictions since one usually guesses the renormalization scale (e.g. usually setting as the one to eliminate large logs and etc.) and varies it over an arbitrary range to ascertain its uncertainty. This ambiguity introduces an important systematic error to pQCD predictions. It has been pointed out that one can use the higher-order $\beta$-terms to achieve an effective value of the strong running coupling $\alpha_s$, and the resultant conformal series is independent to the choice of renormalization scale and thus the conventional renormalization ambiguity is eliminated~\cite{Wu:2013ei, Wu:2014iba, Wu:2019mky}. The principle of maximum conformality (PMC) has been designed for such purpose~\cite{pmc1, pmc2, pmc3, pmc4, pmc5}, which provides a systemic way to eliminate the renormalization scheme-and-scale ambiguities simultaneously. The key idea of PMC is to set the correct momentum flow of the process by absorbing the non-conformal $\beta$-terms that govern the behavior of $\alpha_s$ through the renormalization group equation (RGE). The $\beta_0$-terms in the NLO coefficients can be adopted to set the $\alpha_s$ value, thus in the paper, in addition to the conventional treatment, we shall also adopt the PMC to deal with the $W^+$-boson decays, $W^+ \to B^{(*)}_c+b+\bar{s}+X$.

In the decays, $W^+ \to B^{(*)}_c+b+\bar{s}+X$, the involved hard scales satisfy, $m_W \gg m_b,m_c$, so it is expected that the fragmentation mechanism dominates those decays. The NLO fragmentation functions for a heavy quark to a $B_c$ or $B_c^*$ have recently been given by Ref.\cite{Bcfragnlo}. It is interesting to apply those NLO fragmentation functions to the present processes, and compare the results from the fragmentation approach with those from the fixed-order approach. And in the present paper, besides the fixed-order approach, we shall also adopt the fragmentation approach to do the calculation. In usual cases, because the fragmentation probability of $\bar{b} \to B_c$ is about two orders of magnitude larger than that of $c \to B_c$, the $\bar{b}$ fragmentation is generally more important than the $c$ fragmentation for the $B_c$ meson production. However, in cases with the $W^+$-boson decays, the decay width for the $\bar{b}$ fragmentation shall be depressed due to the small values of the Cabibbo-Kobayashi-Maskawa (CKM) matrix elements $\vert V_{cb} \vert$ and $\vert V_{ub} \vert$; Thus, it provides a good platform to test the fragmentation function for $c \to B_c$.

The paper is organized as follows. In Sec.\ref{secLO}, we briefly present useful formulas at the LO accuracy under the fixed-order approach. In Sec.\ref{secNLO}, we present the formulas to calculate the NLO QCD corrections for the $B^{(*)}_c$ meson production through the $W$-boson decays under the fixed-order approach. In Sec.\ref{secFrag}, we present the useful formulas to calculate the decay widths under the fragmentation approach up to NLO accuracy. In Sec.\ref{secNumer}, numerical results and discussions are presented. Sec.\ref{secCon} is reserved as a summary.

\section{Decay widths at the LO level}
\label{secLO}

According to the NRQCD factorization, the differential decay width for the $B_c$-meson production from the $W^+$-boson decays can be written as
\begin{eqnarray}
&&d\Gamma_{W^+ \to B_c+b+\bar{s}+X}\nonumber \\
&&=\sum_n d\tilde{\Gamma}_{W^+ \to (c\bar{b})[n]+b+\bar{s}+X}\langle{\cal O}^{B_c}(n)\rangle,
\end{eqnarray}
where $d\tilde{\Gamma}_{W^+ \to (c\bar{b})[n]+b+\bar{s}+X}$ denotes the decay width for the production of a perturbative state $(c\bar{b})[n]$ with quantum numbers $[n]$. The long-distance matrix element $\langle{\cal O}^{B_c}(n)\rangle$ is the transition probability for a $(c\bar{b})[n]$ state to the $B_c$ meson. In the lowest-order nonrelativistic approximation, only color-singlet contributions need to be considered, and the long-distance matrix elements for the color-singlet contributions can be determined through potential models.

Practically, we first calculate the decay width for an on-shell $(c\bar{b})$-pair, i.e., $d\Gamma_{W^+ \to (c\bar{b})[n]+b+\bar{s}+X}$. Then the decay width for the $B_c$ meson, i.e., $d\Gamma_{W^+ \to B_c+b+\bar{s}+X}$, can be obtained from $d\Gamma_{W^+ \to (c\bar{b})[n]+b+\bar{s}+X}$ by replacing $\langle{\cal O}^{(c\bar{b})[n]}(n)\rangle$ by $\langle{\cal O}^{B_c}(n)\rangle$.

\begin{figure}[htbp]
\includegraphics[width=0.45\textwidth]{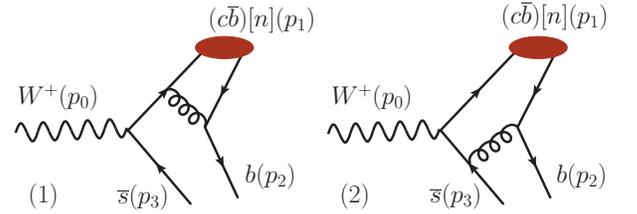}
\caption{The LO Feynman diagrams for $W^+ \to (c\bar{b})[n]+b+\bar{s}$.
 } \label{feylo}
\end{figure}

At the LO level, there are two Feynman diagrams for the $(c\bar{b})[n]$-pair production from the $W^+$-boson decays, which are shown in Fig.\ref{feylo}. The LO amplitude for the $(c\bar{b})[n]$ production can be written as the sum of two terms ($M_{\rm LO}=M_1+M_2$) corresponding to two Feynman diagrams in Fig.\ref{feylo}, and we have
\begin{eqnarray}
iM_1=&&-\frac{i g V_{cs}}{2\sqrt{2}}\frac{-i}{(p_{12}+p_2)^2+i\epsilon}\bar{u}(p_2)(ig_s \gamma^{\mu}T^a) \nonumber \\
&& \cdot \Pi \Lambda_1 (ig_s \gamma_{\mu}T^a) \frac{i}{\slashed{p}_1+\slashed{p}_2-m_c+i \epsilon}  \nonumber \\
&& \cdot \epsilon_{\nu}(p_0) \gamma^{\nu} (1-\gamma_5)v(p_3)\vert_{q=0}, \\
iM_2=&&-\frac{i g V_{cs}}{2\sqrt{2}}\frac{-i}{(p_{12}+p_2)^2+i\epsilon}\bar{u}(p_2)(ig_s \gamma^{\mu}T^a)\Pi \Lambda_1 \nonumber \\
&&  \cdot  \epsilon_{\nu}(p_0)\gamma^{\nu} (1-\gamma_5)\frac{i}{-\slashed{p}_0+\slashed{p}_{11}+i \epsilon} \nonumber \\
&&.(ig_s \gamma_{\mu}T^a)v(p_3)\vert_{q=0},
\end{eqnarray}
where $p_{11}$ and $p_{12}$ are momenta of the $c$ and $\bar{b}$ quarks in $(c\bar{b})[n]$-pair,
\begin{eqnarray}
p_{11}=r_c \,p_1-q,~~ p_{12}=(1-r_c)\,p_1+q,
\end{eqnarray}
where $r_c=m_c/(m_b+m_c)$. $\Pi$ denotes the spin projector, for $^1S_0$ state,
\begin{eqnarray}
\Pi=\frac{-\sqrt{M}}{4m_b m_c}(\slashed{p}_{12}-m_b)\gamma_5(\slashed{p}_{11}+m_c),
\end{eqnarray}
and for $^3S_1$ state,
\begin{eqnarray}
\Pi=\frac{-\sqrt{M}}{4m_b m_c}(\slashed{p}_{12}-m_b)\slashed{\epsilon}(p_1)(\slashed{p}_{11}+m_c).
\end{eqnarray}
$\Lambda_1$ is color-singlet projector, and
\begin{eqnarray}
\Lambda_1=\frac{\textbf{1}}{\sqrt{3}},
\end{eqnarray}
where $\textbf{1}$ denotes the unit matrix of the color $SU(3)$ group.

Using those amplitudes at the LO level, the LO decay width for $(c\bar{b})$-pair production can be calculated through
\begin{eqnarray}
d\Gamma^{(c\bar{b})[n]}_{\rm LO}=\frac{1}{3}\frac{1}{2m_{_W}}\sum \vert M_{\rm LO}\vert^2 d\Phi_3,
\end{eqnarray}
where $\sum$ denotes the sum over the color and spin states of the initial and final particles, $1/3$ comes from the spin average of the initial $W^+$-boson. $d\Phi_3$ denotes the differential phase space at the LO level,
\begin{eqnarray}
d\Phi_3=(2\pi)^d\delta^d\left(p_0-\sum_{f=1}^3 p_f\right)\prod_{f=1}^3 \frac{d^{d-1} \textbf{p}_f}{(2\pi)^{d-1} 2E_f},
\end{eqnarray}
where $d$ stands for the dimension of the space-time. With these formulas, the LO decay width for $W^+ \to (c\bar{b})[n]+b+\bar{s}+X$ can be calculated directly.

\section{The NLO QCD corrections}
\label{secNLO}

The NLO QCD corrections to the decay widths include virtual and real corrections. There are ultraviolet (UV) and infrared (IR) divergences in virtual correction, and IR divergence in real correction. We adopt the conventional dimensional regularization approach with $d=4-2\epsilon$ to regulate these divergences. Then the UV and IR divergences appear as pole terms in $1/\epsilon$. We shall sketch the calculations for the virtual and real corrections in the following subsections.

\subsection{The virtual correction}

\begin{figure}[htbp]
\includegraphics[width=0.45\textwidth]{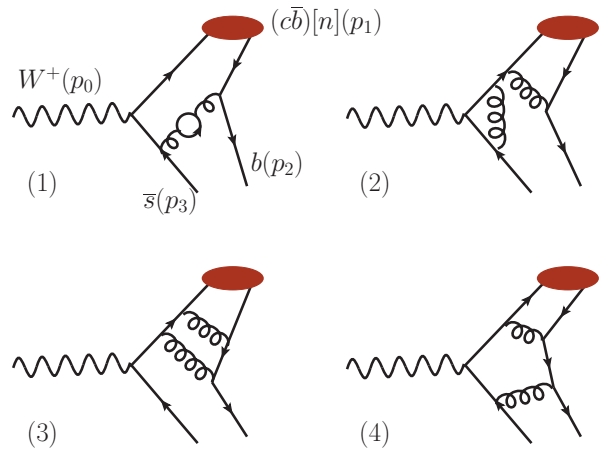}
\caption{Four typical one-loop Feynman diagrams for $W^+ \to c\bar{b}[n]+b+\bar{s}$.
 } \label{feyvir}
\end{figure}

Four typical one-loop diagrams are shown in Fig.\ref{feyvir}. The NLO virtual corrections come from the interference of those one-loop Feynman diagrams with the LO Feynman diagrams shown in Fig.\ref{feylo}.

The virtual corrections can be calculated through
\begin{equation}
d\Gamma_{\rm Virtual}^{(c\bar{b})[n]}=\frac{1}{3}\frac{1}{2m_{_W}}\sum 2 {\rm Re}\left(M^*_{\rm LO} M_{\rm Virtual} \right)d\Phi_3,
\end{equation}
where $M_{\rm Virtual}$ denotes the amplitude for the virtual corrections, $d\Phi_3$ is the LO differential phase space.

As a subtle point, there are Coulomb divergences in the hard part of the NLO amplitudes by using the traditional matching procedure. One may observe hat they appear in both the virtual corrections to the $(c\bar{b})[n]$ production and the virtual corrections to the long-distance matrix element $\langle{\cal O}^{(c\bar{b})[n]}(n)\rangle$, which shall be canceled by each other. As a result, no Coulomb divergence appears in the resultant pQCD series. In dimensional regularization, there is a simpler way to extract the NRQCD short-distance coefficients using the method of regions~\cite{region}. In this method, one can calculate the hard region contributions directly by expanding the relative momentum of the $(c\bar{b})[n]$ pair before carrying out the loop integration (More explicitly, under the present lowest-order nonrelativistic approximation, one just needs to set $q=0$ before the loop integration). Then the Coulomb divergences, which are power IR divergences, vanish in dimensional regularization. We adopt this new treatment, and the Coulomb divergences shall not appear in our present calculation.

There are UV and IR divergences in the loop-diagram contributions. The IR divergences in the virtual correction shall be canceled by the IR divergences in the real correction. The UV divergences should be removed through renormalization. We carry out the renormalization using counterterm approach, where the decay widths are calculated in terms of the renormalized quark mass $m$, the renormalized quark field $\Psi_r$, the renormalized gluon field $A_r^{\mu}$, and the renormalized coupling constant $g_s$. The relations between the renormalized quantities and their corresponding bare quantities are
\begin{eqnarray}
m_0=Z_m m,~~\Psi_0=\sqrt{Z_2} \Psi_r,\nonumber \\
A_0^{\mu}=\sqrt{Z_3}A_r^{\mu},~~~~g_s^0=Z_g g_s,
\end{eqnarray}
where $Z_i=1+\delta Z_i$ with $i=m,2,3,g$ are renormalization constants, and they are fixed by the renormalization scheme. The renormalization scheme is adopted as follows: The renormalization of the heavy quark mass, the heavy quark field, the light quark field and the gluon field are performed in the on-shell scheme, whereas the renormalization of the strong coupling constant is performed in the $\overline{\rm MS}$ scheme. The quantities $\delta Z_i$ can be calculated and they are
\begin{eqnarray}
\delta Z^{\rm OS}_{m,Q}&=&-3~C_F \frac{\alpha_s}{4\pi}\left[\frac{1}{\epsilon_{UV}}- \gamma_E+
 {\rm ln}\frac{4\pi \mu_R^2}{m_Q^2}+\frac{4}{3}\right],\nonumber\\
 \delta Z^{\rm OS}_{2,Q}&=&-C_F \frac{\alpha_s}{4\pi}\left[\frac{1}{\epsilon_{UV}}+ \frac{2}{\epsilon_{IR}}-3~\gamma_E+3~ {\rm ln}\frac{4\pi \mu_R^2}{m_Q^2}+4\right], \nonumber\\
  \delta Z^{\rm OS}_{2,q}&=&-C_F \frac{\alpha_s}{4\pi}\left[\frac{1}{\epsilon_{UV}}- \frac{1}{\epsilon_{IR}}\right], \nonumber\\
 \delta Z^{OS}_3&=&\frac{\alpha_s}{4\pi}\left[(\beta'_0-2C_A) \left(\frac{1}{\epsilon_{UV}}-\frac{1}{\epsilon_{IR}}\right) \right. \nonumber\\
 &&\left.-\frac{4}{3}T_F \left(\frac{1}{\epsilon_{UV}}-\gamma_E + {\rm ln}\frac{4\pi \mu_R^2}{m_c^2}\right)\right. \nonumber\\
 &&\left.-\frac{4}{3}T_F \left(\frac{1}{\epsilon_{UV}}-\gamma_E + {\rm ln}\frac{4\pi \mu_R^2}{m_b^2}\right)\right], \nonumber\\
 \delta Z^{\overline{\rm MS}}_g&=&- \frac{\beta_0}{2}\frac{\alpha_s}{4\pi}\left[\frac{1}{\epsilon_{UV}}- \gamma_E+ {\rm ln}~(4\pi) \right], \nonumber
\end{eqnarray}
where $Q (=c,b)$ in the subscripts denotes a heavy quark, and $q(=s)$ denotes a light quark. $\mu_R$ is the renormalization scale, $\gamma_E$ is the Euler constant. For QCD, $C_A=3$, $C_F=4/3$ and $T_F=1/2$. $\beta_0=11C_A/3-4T_F n_f/3$ is the one-loop coefficient of the QCD $\beta$ function, in which $n_f$ is the number of active quark flavors. $\beta'_0=11C_A/3-4T_F n_{lf}/3$ and $n_{lf}=3$ is the number light-quark flavors.

\subsection{The real correction}

\begin{figure}[htbp]
\includegraphics[width=0.45\textwidth]{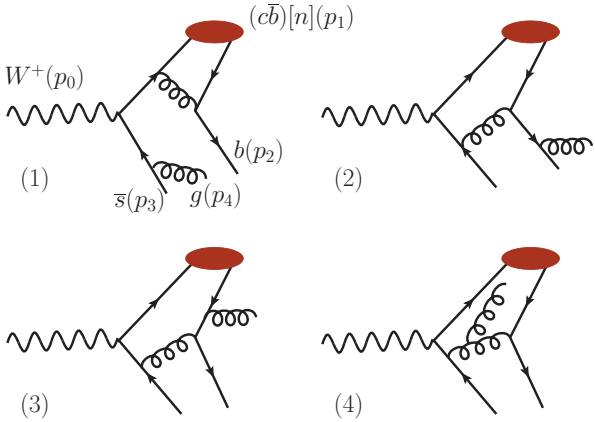}
\caption{Four typical real-correction Feynman diagrams for $W^+ \to (c\bar{b})[n]+b+\bar{s}$.
 } \label{feyreal}
\end{figure}

The real corrections come from the decay process $W^+(p_0) \to (c\bar{b})[n](p_1)+b(p_2)+\bar{s}(p_3)+g(p_4)$. The Feynman diagrams for the real corrections can be obtained through the LO Feynman diagrams by adding an additional gluon in the final state. Typical real correction Feynman diagrams are shown in Fig.\ref{feyreal}. The real correction can be calculated through
\begin{eqnarray}
d\Gamma_{\rm Real}^{(c\bar{b})[n]}=\frac{1}{3}\frac{1}{2m_{_W}}\sum \vert M_{\rm Real} \vert^2 d\Phi_4,
\end{eqnarray}
where $d\Phi_4$ denotes the differential phase space for the real corrections, and
\begin{eqnarray}
d\Phi_4=(2\pi)^d\delta^d\left(p_0-\sum_{f=1}^4 p_f\right)\prod_{f=1}^4 \frac{d^{d-1} \textbf{p}_f}{(2\pi)^{d-1} 2E_f}.
\end{eqnarray}

There are IR divergences in the real correction, which come from the phase-space integration. However, integrating the squared amplitude directly over the phase space in $d$ dimensions is too difficult to be practical. In order to isolate the divergent and finite terms, we adopt the two-cutoff phase-space slicing method~\cite{twocutoff} to calculate the real correction. Following this method, the phase space for the real correction is decomposed into three regions by introducing two small cutoffs, $\delta_s$ and $\delta_c$, which should satisfy the requirement $\delta_c \ll \delta_s$~\cite{twocutoff}. The three regions are soft region (${\rm S}$) with $E_4 \leq m_{_W}\,\delta_s /2$, hard-collinear region (${\rm HC}$) with $E_4>m_{_W}\,\delta_s /2$ and $(p_3+p_4)^2\leq \delta_c m_{_W}^2$, and hard-non-collinear region (${\rm H}\bar{\rm C}$) with  $E_4>m_{_W}\,\delta_s /2$ and $(p_3+p_4)^2> \delta_c m_{_W}^2$, where $E_4$ is defined in the rest frame of the initial $W^+$ boson. The IR finite hard-non-collinear part can be calculated numerically in four dimensions.

Applying the soft approximation to the soft part (e.g. all the terms of order $\delta_s$ are neglected), we obtain
\begin{eqnarray}
d\Gamma^{(c\bar{b})[n]}_{\rm S}=&&d\Gamma^{(c\bar{b})[n]}_{\rm LO}\left[\frac{C_F \alpha_s}{2\pi}\frac{\Gamma(1-\epsilon)}{\Gamma(1-2 \epsilon)}\left(\frac{4\pi\mu_R^2}{m_{_W}^2}\right)^{\epsilon}\right] \nonumber \\
&&\cdot \left\{\frac{1}{\epsilon^2}\right. +\frac{1}{\epsilon}\left[1-2\,{\rm ln}\delta_s -{\rm ln}\left(\frac{(1-\beta_b\,{\rm cos}\theta)^2}{1-\beta_b^2} \right)\right] \nonumber \\
&&-2\,{\rm ln}\delta_s+2\,{\rm ln}^2\delta_s+2\,{\rm ln}\delta_s \,{\rm ln} \left(\frac{(1-\beta_b\,{\rm cos}\theta)^2}{1-\beta_b^2} \right)\nonumber \\
&&+\frac{1}{\beta_b}{\rm ln}\left(\frac{1+\beta_b}{1-\beta_b} \right)+{\rm ln}^2\left(\frac{1-\beta_b}{1-\beta_b \,{\rm cos}\theta}\right) \nonumber \\
&&-\frac{1}{2}\,{\rm ln}^2\left(\frac{1+\beta_b}{1-\beta_b} \right)+2\,{\rm Li}_2\left(-\frac{\beta_b(1-{\rm cos}\theta)}{1-\beta_b} \right)\nonumber \\
&&\left. -2\,{\rm Li}_2\left(-\frac{\beta_b(1+{\rm cos}\theta)}{1-\beta_b\,{\rm cos}\theta} \right)\right \},
\end{eqnarray}
where $\beta_b=\sqrt{1-m_b^2/E_2^2}$, and $E_2$ is defined in the rest frame of the $W^+$ boson. $\theta$ is the angle between $\textbf{p}_2$ and $\textbf{p}_3$ in the rest frame of the $W^+$ boson.

Applying the collinear approximation (e.g. terms of order $\delta_c$ are neglected), we obtain
\begin{eqnarray}
d\Gamma^{(c\bar{b})[n]}_{\rm HC}=&&d\Gamma^{(c\bar{b})[n]}_{\rm LO}\left[\frac{C_F \alpha_s}{2\pi}\frac{\Gamma(1-\epsilon)}{\Gamma(1-2 \epsilon)}\left(\frac{4\pi\mu_R^2}{m_{_W}^2}\right)^{\epsilon}\right]\nonumber \\
&&\times \left(\frac{A_1^{q\to qg}}{\epsilon}+A_0^{q\to qg} \right),
\end{eqnarray}
where
\begin{eqnarray}
A_1^{q\to qg} &=& 3/2+2{\rm ln}\delta'_s , \\
A_0^{q\to qg} &=& 7/2-\pi^2/3-{\rm ln}^2\delta'_s -{\rm ln}\delta_c(3/2+2{\rm ln}\delta'_s),
\end{eqnarray}
and $\delta'_s=m_{_W}^2 \delta_s/[m_{_W}^2-(p_1+p_2)^2]$.

Summing up three parts from the soft, hard-collinear and hard-non-collinear regions, we obtain the required real correction. Separate contributions from three regions depend on one or both of the two cutoff parameters $\delta_s$ and $\delta_c$. However, the sum of those three contributions should be independent to the choices of $\delta_s$ and $\delta_c$. The verification of this cut-off independence provides an important check for the correctness of the numerical calculation. We have checked this independence and have indeed found that the results are independent of the $\delta_s$ and $\delta_c$ by varying $\delta_s$ from $10^{-3}$ to $10^{-7}$ with $\delta_c=\delta_s/50$. For clarity, we fix $\delta_s=10^{-5}$ and $\delta_c=2\times 10^{-7}$ to do the following numerical calculation.

Total NLO corrections can be obtained by summing up virtual and real corrections. The UV and IR divergences are exactly canceled after summing the real and virtual corrections, and the finite NLO corrections are obtained. Then the decay widths $\Gamma_{W^+\to B_c+b+\bar{s}+X}$ and $\Gamma_{W^+\to B^*_c+b+\bar{s}+X}$ can be derived from $\Gamma_{W^+\to (c\bar{b})[^1S_0]+b+\bar{s}+X}$ and $\Gamma_{W^+\to (c\bar{b})[^3S_1]+b+\bar{s}+X}$ by multiplying a factor $\langle {\cal O}^{B_c(B_c^*)}(n) \rangle/\langle {\cal O}^{(c\bar{b})[n]}(n) \rangle \approx \vert R_S(0)\vert^2/4\pi$, where $n=\,^1S_0$ for $B_c$ and $n= \,^3S_1$ for $B^*_c$, respectively. $R_S(0)$ denotes the radial wave function at the origin of the $B_c(B^*_c)$ meson.

In the calculation, we adopt the FeynArts package~\cite{feynarts} to generate the Feynman diagrams and the corresponding amplitudes, and the FeynCalc package~\cite{feyncalc1, feyncalc2} to carry out the Dirac and color traces. Then we use the \$Apart package~\cite{apart} and the FIRE package~\cite{fire} to do partial fraction and integration-by-parts (IBP) reduction of the loop integrals. After the IBP reduction, there are only few master integrals (e.g. $A_0$, $B_0$, $C_0$, and $D_0$ functions) need to be calculated, which shall be dealt with by using the LoopTools package~\cite{looptools}. Numerical phase-space integrations are carried out by the VEGAS program~\cite{vegas}.

\section{Decay widths under the fragmentation approach}
\label{secFrag}

We take the process $W^+ \to B_c+X$ as an example to illustrate the calculation under the fragmentation approach. The formulas for the $B_c^*$ production are similar to the $B_c$ case.

The differential decay width for $W^+ \to B_c+X$ under the fragmentation approach can be written as
\begin{eqnarray}
\frac{d\Gamma_{W^+\to B_c+X}}{dz}=&& \sum_i \int^1_z \frac{dy}{y}\frac{d\hat{\Gamma}_{W^+ \to i+X}(y,\mu_F)}{dy}\cdot \nonumber \\
&& D_{i \to B_c}(z/y,\mu_F),\label{eq.frag}
\end{eqnarray}
where $d\hat{\Gamma}_{W^+ \to i+X}(y,\mu_F)$ denotes the decay width (coefficient function) for a $W^+$ to a parton $i$\footnote{Due to the coefficient function $d\hat{\Gamma}_{W^+ \to i+X}(y,\mu_F)$ is IR safe, the heavy-quark mass $m_Q$ in the coefficient function can be approximately set to $0$, and this approximation brings only a small error of ${\cal O}(m_Q^2/m_{_W}^2)$. In the following fragmentation calculations, we shall adopt this approximation for simplicity. The neglected higher-power terms will be included in the results by combining the fixed-order and fragmentation approaches.}, $D_{i \to B_c}(z/y,\mu_F)$ denotes the fragmentation function for a parton $i$ into a $B_c$, and $\mu_F$ is the factorization scale which separates the energy scales of two parts.

For comparison, we adopt several strategies to obtain the fragmentation predictions. More details about those strategies can be found in Refs.\cite{Bcfragnlo, jpsifragnlo}. For convenience, we denote them as ``Frag, LO", ``Frag, NLO" and ``Frag, NLO+NLL", respectively. For the case of ``Frag, LO",
\begin{eqnarray}
\frac{d\Gamma^{\rm Frag, LO}_{W^+ \to B_c +X}}{dz}=&& \int^1_z \frac{dy}{y}\frac{d\hat{\Gamma}^{\rm LO}_{W^+ \to c+\bar{s}}(y,\mu_F)}{dy} \nonumber \\
&& \cdot D^{\rm LO}_{c \to B_c}(z/y,\mu_F)\nonumber \\
=&&\Gamma^{\rm LO}_{W^+ \to c+\bar{s}}\cdot D^{\rm LO}_{c \to B_c}(z,\mu_F),
\end{eqnarray}
where $\Gamma^{\rm LO}_{W^+ \to c+\bar{s}}$ is the LO decay width for $W^+ \to c+\bar{s}$ and $D^{\rm LO}_{c \to B_c}(z,\mu_F)$ is the LO fragmentation function. In the calculation, the factorization and renormalization scales are set as $\mu_F=2m_b+m_c$ and $\mu_R=2m_b$. 

For the case of ``Frag, NLO",
\begin{eqnarray}
\frac{d\Gamma^{\rm Frag, NLO}_{W^+ \to B_c +X}}{dz}=&& \int^1_z \frac{dy}{y}\frac{d\hat{\Gamma}^{\rm NLO}_{W^+ \to c+X}(y,\mu_F)}{dy} \nonumber \\
&& \cdot D^{\rm NLO}_{c \to B_c}(z/y,\mu_F),   \label{fragnlo}
\end{eqnarray}
where the NLO fragmentation function $D^{\rm NLO}_{c \to B_c}(z,\mu_F)$ can be found in Ref.\cite{Bcfragnlo}. In the calculation, the factorization and renormalization scales are set as $\mu_F=2m_b+m_c$ and $\mu_R=2m_b$, and the factorization scheme is chosen as the $\overline{\rm MS}$ scheme. 

For the case of ``Frag, NLO+NLL",
\begin{eqnarray}
\frac{d\Gamma^{\rm Frag, NLO+NLL}_{W^+ \to B_c +X}}{dz}=&& \int^1_z \frac{dy}{y}\frac{d\hat{\Gamma}^{\rm NLO}_{W^+ \to c+X}(y,\mu_F)}{dy} \nonumber \\
&& \cdot D^{\rm NLO+NLL}_{c \to B_c}(z/y,\mu_F),  \label{fragnlonll}
\end{eqnarray}
where the upper factorization and renormalization scales in the coefficient function $d\hat{\Gamma}^{\rm NLO}_{W^+ \to c+X}(y,\mu_F)/dy$ are set as $\mu_F=\mu_R=m_{_W}$, so as to avoid the large logarithms of $\mu_F^2/m_{_W}^2$ or $\mu_R^2/m_{_W}^2$ appear in the coefficient function. The fragmentation function $D^{\rm NLO+NLL}_{c \to B_c}(z,\mu_F=m_{_W})$ is obtained through solving the Dokshitzer-Gribov-Lipatov-Altarelli-Parisi (DGLAP) evolution equation~\cite{dglap1, dglap2, dglap3} with NLO splitting function for $c \to c$~\cite{nlospfun1, nlospfun2, nlospfun3}, where the NLO fragmentation function $D^{\rm NLO}_{c \to B_c}(z,\mu_{F0}=2m_b+m_c)$ with the lower factorization and renormalization scales $\mu_{F0}=2m_b+m_c$ and $\mu_{R0}=2m_b$ is used as the boundary condition. To solve the DGLAP evolution equation, the Mellin transformation method is adopted, and the related formulas for the Mellin transformation method can be found in Ref.\cite{MELE} \footnote{In Ref.\cite{MELE}, the authors adopted an approximation that the sigularity of the NLO splitting function $P_{Q\to Q}(z)$ at $z=1$ is regularized by an overall ``+" prescription. Then, $\int_0^1 P_{c \to c}(z) \,dz=0$. This approximation is good, since the probability to produce an additional heavy quark pair in the evolution process is very low at the energy of $m_{_W}$. For simplicity, we shall adopt this approximation in the present paper.}.

\section{Numerical results and discussions}
\label{secNumer}

To do the numerical calculation, the input parameters are taken as follows:
\begin{eqnarray}
&& m_b=4.9\,{\rm GeV},\; m_c=1.5\,{\rm GeV},\;m_{_W}=80.4\,{\rm GeV},\nonumber \\
&& G_F=1.166\times10^{-5}\,{\rm GeV}^{-2},\; \vert V_{cs}\vert=1,\nonumber \\
&& \vert R_S(0)\vert^2=1.642\,{\rm GeV}^{3},
\end{eqnarray}
where $G_F$ is the Fermi coupling constant. The input value for $\vert R_S(0)\vert^2$ is taken from the potential-model calculation~\cite{pot}. For the strong coupling constant, we use the two-loop formula
\begin{displaymath}
\alpha_s(\mu_R)=\frac{4\pi}{\beta_0{\rm ln}(\mu_R^2/\Lambda^2_{QCD})}\left[ 1-\frac{\beta_1{\rm ln}\,{\rm ln}(\mu_R^2/\Lambda^2_{QCD})}{\beta_0^2\,{\rm ln}(\mu_R^2/\Lambda^2_{QCD})}\right],
\end{displaymath}
where $\beta_1=34\,C_A^2/3-4\,T_F \,C_F n_f-20\,T_F\, C_A n_f/3$ is the two-loop coefficient of the QCD $\beta$-function. According to $\alpha_s(m_{_Z})=0.1185$~\cite{pdg}, we obtain $\Lambda^{n_f=5}_{\rm QCD}=0.233\,{\rm GeV}$ and $\Lambda^{n_f=4}_{\rm QCD}=0.337\,{\rm GeV}$.

\subsection{Basic results}

\begin{table}[h]
\begin{tabular}{c c c c c}
\hline
  & $\Gamma_{\rm LO}$(keV) & $\Gamma^{\rm Cor.}_{\rm NLO}$(keV) & $\Gamma_{\rm NLO}$(keV) & $\Gamma^{\rm Cor.}_{\rm NLO}/\Gamma_{\rm LO}$ \\
\hline
$\mu_R=2m_b$    & 2.89 & 2.00 & 4.89  & 0.69  \\
$\mu_R=m_{_W}$  & 1.30 & 1.42 & 2.72  & 1.09 \\
\hline
\end{tabular}
\caption{The total decay widths of $W^+\to B_c+b+\bar{s}+X$ up to NLO level under the fixed-order approach. Two typical renormalization scales are adopted.}
\label{tb.wbcwidth}
\end{table}

\begin{table}[h]
\begin{tabular}{c c c c c }
\hline
  &  $\Gamma_{\rm LO}$(keV)& $\Gamma^{\rm Cor.}_{\rm NLO}$(keV) & $\Gamma_{\rm NLO}$(keV) & $\Gamma^{\rm Cor.}_{\rm NLO}/\Gamma_{\rm LO}$ \\
\hline
$\mu_R = 2m_b$ &   2.48 & 1.07 & 3.55  & 0.43  \\
$\mu_R = m_{_W}$ & 1.12 & 1.03 & 2.15  & 0.92 \\
\hline
\end{tabular}
\caption{The total decay widths of $W^+\to B^*_c+b+\bar{s}+X$ up to NLO level under the fixed-order approach. Two typical renormalization scales are adopted.}
\label{tb.wbc*width}
\end{table}

The decay widths for $W^+\to B_c+b+\bar{s}+X$ and $W^+\to B^*_c+b+\bar{s}+X$ under the fixed-order approach are given in Tables \ref{tb.wbcwidth} and \ref{tb.wbc*width}, where $\Gamma_{\rm NLO}=\Gamma_{\rm LO}+\Gamma^{\rm Cor.}_{\rm NLO}$ and $\Gamma^{\rm Cor.}_{\rm NLO}$ denotes the NLO corrections, $\Gamma^{\rm Cor.}_{\rm NLO}=\Gamma_{\rm virtual}+\Gamma_{\rm Real}$. In the calculation, we take two typical energy scales ($2m_b$ and $m_{_W}$) as the renormalization scale, and we have $\alpha_s(2m_b)=0.180$ and $\alpha_s(m_W)=0.121$. Tables \ref{tb.wbcwidth} and \ref{tb.wbc*width} show that the NLO corrections are significant. After including the NLO corrections, the total decay width for $W^+\to B_c(B^*_c)+b+\bar{s}+X$ is increased by $69\%$ $(43\%)$ for $\mu_R=2m_b$ and $109\%$ $(92\%)$ for $\mu_R=m_{_W}$.

\begin{figure}[htbp]
\includegraphics[width=0.5\textwidth]{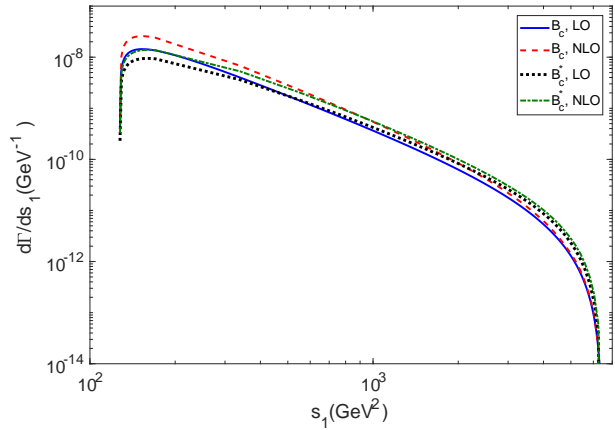}
\caption{Differential decay widths $d\Gamma/ds_1$ for $W^+\to B_c (B^*_c)+b+\bar{s}+X$ under the fixed-order approach. $\mu_R=2m_b$.} \label{gammas1}
\end{figure}

\begin{figure}[htbp]
\includegraphics[width=0.5\textwidth]{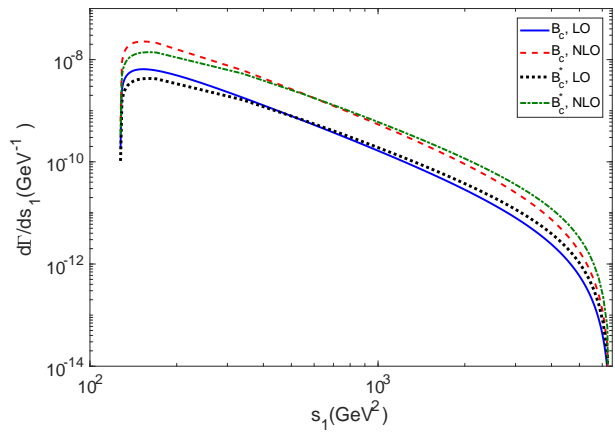}
\caption{Differential decay widths $d\Gamma/ds_1$ for $W^+\to B_c (B^*_c)+b+\bar{s}+X$ under the fixed-order approach. $\mu_R=m_W$.} \label{gammas2}
\end{figure}

The momentum of the produced $b$-quark jet can be measured using vertex tagging technology, so the invariant mass of the $B^{(*)}_c$ and $b$ quark in the final state can be determined experimentally. We present the differential decay widths $d\Gamma/ds_1$ for $W^+\to B_c (B^*_c)+b+\bar{s}+X$ in Figs. \ref{gammas1} and \ref{gammas2}, where $\mu_R=2m_b$ and $\mu_R=m_W$, respectively. Here $s_1\equiv (p_1+p_2)^2$. Figs. \ref{gammas1} and \ref{gammas2} show that there is a peak near the minimum value of $s_1$, indicating the dominant contributions of the decay processes come from the phase-space region near the threshold of producing the $B_c(B_c^*)$ and $b$ quark. This property may be helpful to distinguish the $B_c$ or $B^*_c$ mesons produced through the $W^+$-boson decays from the those produced through other production mechanisms at the high-energy colliders.

Tables \ref{tb.wbcwidth} and \ref{tb.wbc*width} show that after including the NLO corrections, the renormalization scale dependence is softened. However, such scale dependence is still very large, e.g. $W^+\to B_c(B^*_c)+b+\bar{s}+X$, the NLO total decay width decreases by $44\%$ $(39\%)$, when $\mu_R$ changes from $2m_b$ to $m_{_W}$.

As mentioned in the Introduction, the PMC scale-setting approach provides a way to eliminate the renormalization scale ambiguity~\cite{pmc1, pmc2, pmc3, pmc4, pmc5}. As an attempt of showing how the PMC affects the decay width, we present the PMC predictions in the following.

To apply the PMC, we first schematically rewritten the NLO decay width as
\begin{eqnarray}
\Gamma &=& A\, \alpha^2_s(\mu_R) \left[ 1+ (a+b\, n_f)\frac{\alpha_s(\mu_R)}{\pi}\right]\nonumber \\
&=&A\, \alpha^2_s(\mu_R) \bigg \lbrace 1+ \left[-\frac{3b}{2}\beta_0+\left(a+\frac{33b}{2}\right)\right]\frac{\alpha_s(\mu_R)}{\pi}\bigg\rbrace,
\end{eqnarray}
Using the RGE, the non-conformal term ($-\frac{3b}{2}\beta_0$) can be adopted to fix the strong running coupling. A PMC scale $\mu^{\rm PMC}$ is then determined, which corresponds to the (correct) typical momentum flow of the process. Then, following the standard PMC procedures, the NLO decay width changes to
\begin{equation}
\Gamma_{\rm NLO}^{\rm PMC}= A\, \alpha_s(\mu^{\rm PMC})^2 \left[ 1+ \left(a+\frac{33b}{2}\right)\alpha_s(\mu^{\rm PMC})\right],
\end{equation}
where $\mu^{\rm PMC}=\mu_R\,e^{3b/2}$. It is interesting to find that the PMC scale $\mu^{\rm PMC}$ is independent to any choice of renormalization scale $\mu_R$, e.g. $\mu^{\rm PMC}\equiv 6.67$ GeV for $B_c$ and $\mu^{\rm PMC}\equiv 7.17$ GeV for $B^*_c$, thus the conventional renormalization scale ambiguity is really eliminated. The PMC scales are closer to $\mu_R=2m_b$ than $\mu_R=m_{_W}$, the conditions for the total decay widths are similar, thus the usual guessing choice of $\mu_R=2m_b$ is more reasonable for conventional prediction. Thus in the following analysis, we fix $\mu_R=2m_b$ for predictions when use conventional pQCD series.

\begin{table}[htb]
\begin{tabular}{c c c c c}
\hline
  &   $\Gamma_{\rm LO}$ (keV) & $\Gamma_{\rm NLO,Cor.}$ (keV) & $\Gamma_{\rm NLO}$ (keV) & $\Gamma_{\rm NLO,Cor.}/\Gamma_{\rm LO}$  \\
\hline
$B_c$   &   3.53  & 2.05 & 5.58 & 0.58   \\
$B_c^*$ &   2.92  & 0.92 & 3.84 & 0.32   \\
\hline
\end{tabular}
\caption{Total decay widths of $W^+\to B_c (B^*_c)+b+\bar{s}+X$ up to NLO accuracy under the PMC scale-setting approach.}
\label{tb.wbcpmc}
\end{table}

Numerical results for the total decay widths of $W^+\to B_c (B^*_c)+b+\bar{s}+X$ up to NLO accuracy under the PMC are shown in Table \ref{tb.wbcpmc}. After applying the PMC scale-setting approach, the convergence is slightly better than conventional series, e.g. after including the NLO QCD corrections, the decay width for $W^+\to B_c(B^*_c)+b+\bar{s}+X$ is increased by $58\%$ $(32\%)$.

\subsection{Comparison of the decay widths under the fixed-order and fragmentation approaches}

\begin{figure}[htbp]
\includegraphics[width=0.5\textwidth]{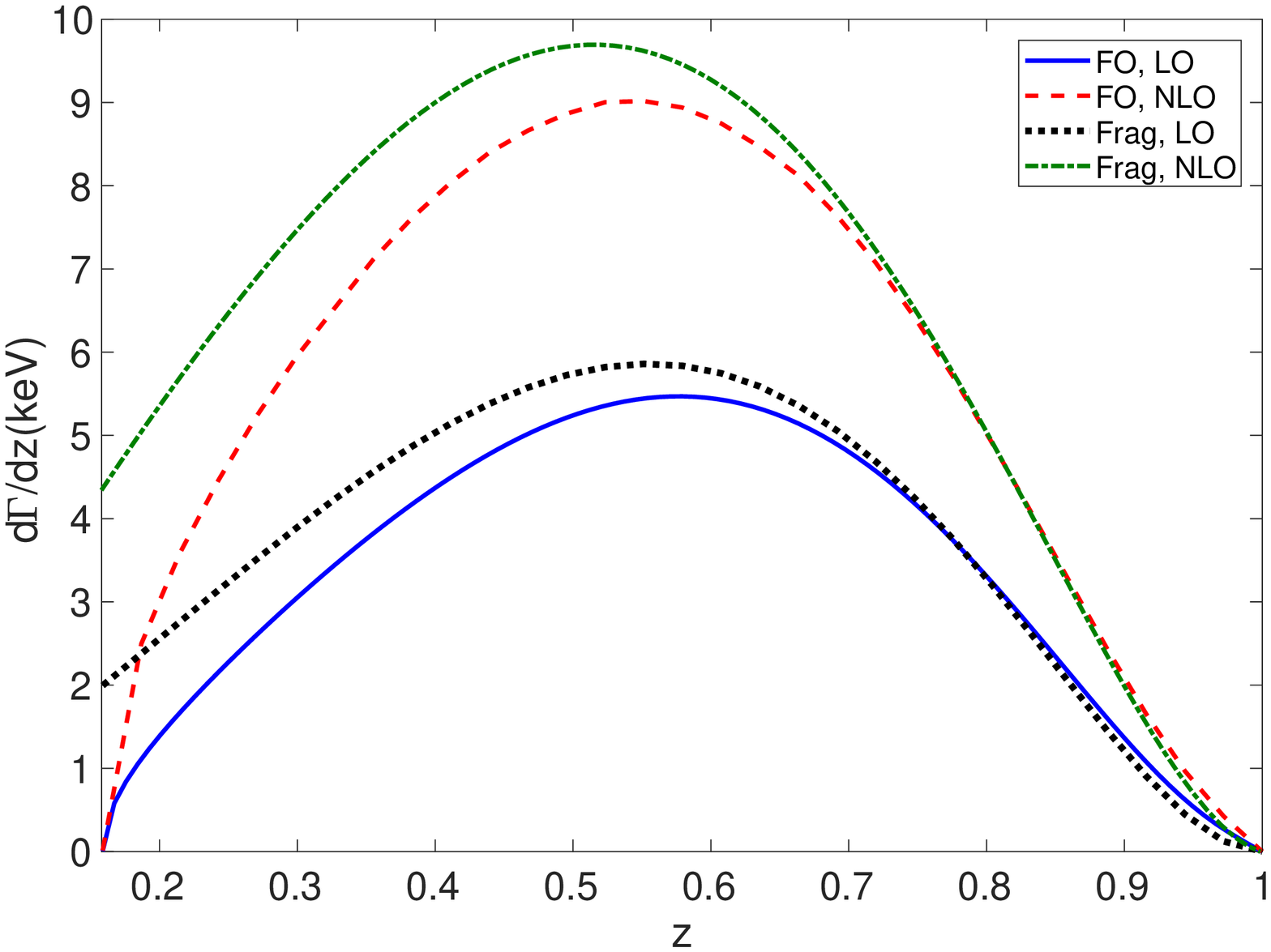}
\caption{The differential width $d\Gamma/dz$ for $W^+\to B_c+b+\bar{s}+X$ under the fixed-order (FO) and fragentation (Frag) approaches.} \label{gammaz}
\end{figure}

\begin{figure}[htbp]
\includegraphics[width=0.5\textwidth]{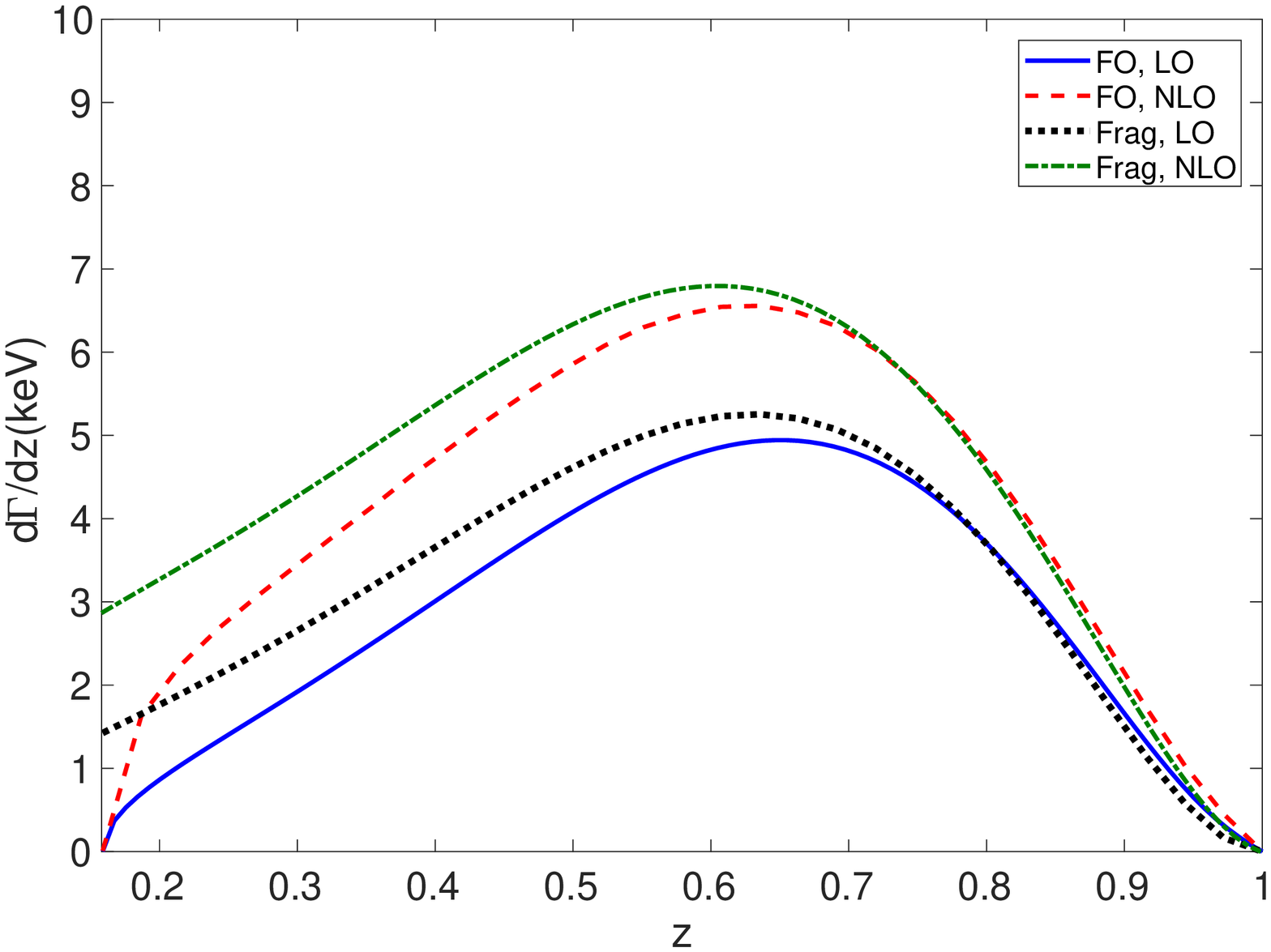}
\caption{The differential width $d\Gamma/dz$ for $W^+\to B^*_c+b+\bar{s}+X$ under the fixed-order (FO) and fragentation (Frag) approaches.} \label{gammaz3}
\end{figure}

It is interesting to know the differential distributions of those decay processes. We define the energy fraction $z\equiv E_1/E_1^{\rm max}$, where $E_1$ and $E_1^{\rm max}$ are the energy and the maximum energy of the $B_c(B_c^*)$ meson in the rest frame of the initial $W^+$ boson. The differential decay widths $d\Gamma/dz$ for $W^+\to B^*_c+b+\bar{s}+X$ are presented in Figs. \ref{gammaz} and \ref{gammaz3}. In addition to the fixed-order results, we also present the results from the fragmentation approach up to NLO level. Here, in order to know whether the fragmentation mechanism dominates the decay processes, we do not resum the leading logarithms of $m_{Q}^2/m_{_W}^2$, i.e., the coefficient function and the fragmentation functions are both calculated at the LO or NLO level without the DGLAP evolution. The factorization and renormalization scales are set as $2m_b+m_c$ and $2m_b$ respectively in the fragmentation calculation, and the renormalization scale is set as $2m_b$ in the fixed-order calculation.

\begin{figure}[htbp]
\includegraphics[width=0.5\textwidth]{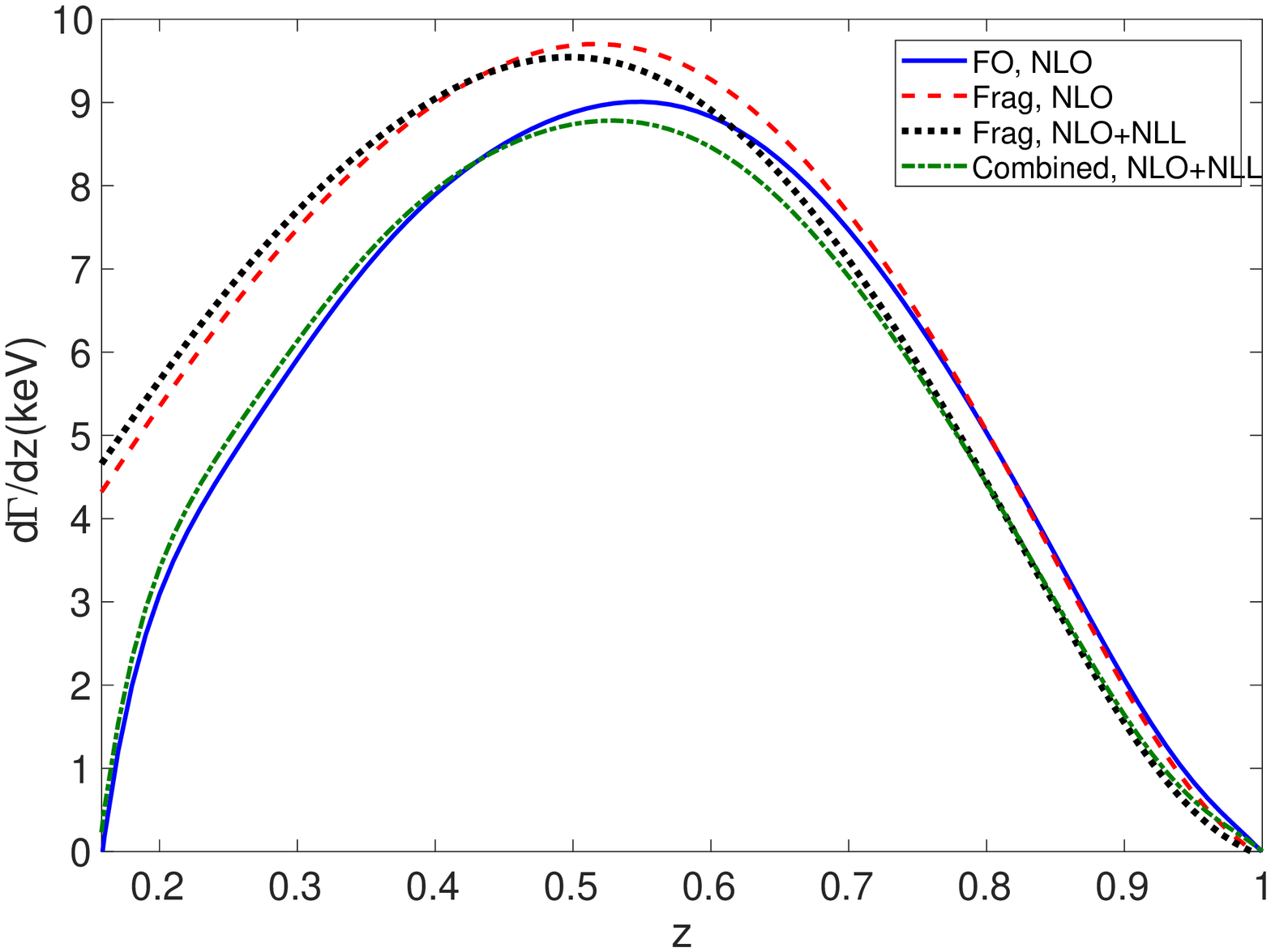}
\caption{The NLO differential decay width $d\Gamma/dz$ for $W^+\to B_c+b+\bar{s}+X$ under the fixed-order approach (FO), fragmentation approach (Frag) and the combination of two approaches (Combined), respectively.} \label{gammaz1}
\end{figure}

\begin{figure}[htbp]
\includegraphics[width=0.5\textwidth]{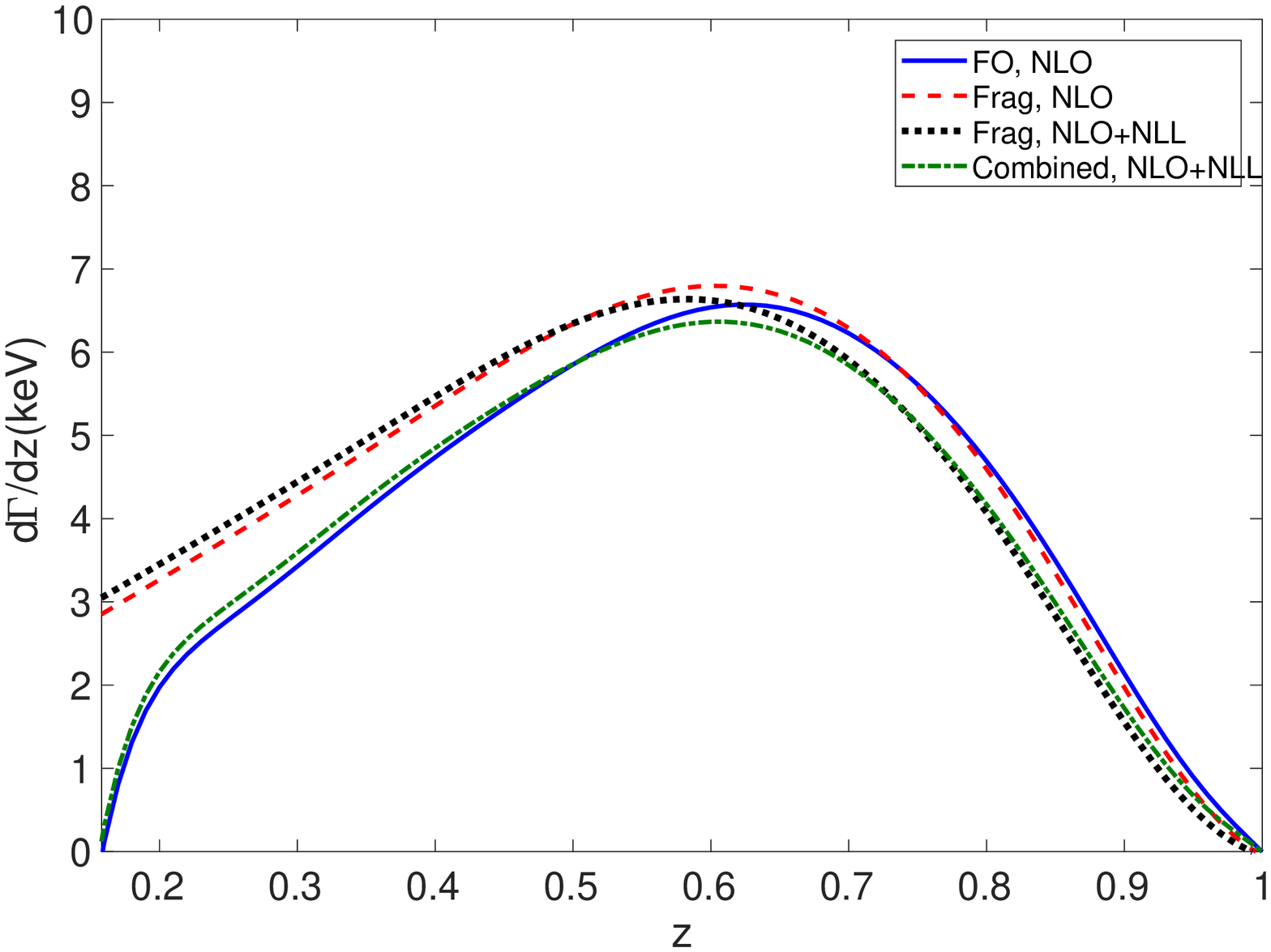}
\caption{The NLO differential decay width $d\Gamma/dz$ for $W^+\to B^*_c+b+\bar{s}+X$ under the fixed-order approach (FO), fragentation  approach (Frag) and the combination of two approaches (Combined), respectively.} \label{gammaz31}
\end{figure}

\begin{table}[h]
\begin{tabular}{c c c}
\hline
    & ~~Frag, NLO~~ & ~~Frag, NLO+NLL~~   \\
\hline
$B_c$    & 5.82  &  5.71   \\
$B_c^*$  & 4.15  &  4.07   \\
\hline
\end{tabular}
\caption{The total decay widths (in unit: keV) for $W^+\to B_c (B^*_c)+b+\bar{s}+X$ under the fragmentation approach.}
\label{tb.wbcFrag}
\end{table}

Figs. \ref{gammaz} and \ref{gammaz3} show that the fragmentation mechanism dominates the decay processes $W^+\to B_c (B^*_c)+b+\bar{s}+X$, since the fixed-order and fragmentation shapes are close at the LO and NLO levels. Differences appear in small $z$ region, indicating in this $z$-region, the non-fragmentation terms become important. The neglected logarithms of $m_{Q}^2/m_{_W}^2$ may give sizable contributions, which can be resumed in fragmentation approach by using the DGLAP evolution equation. More explicitly, the NLO fragmentation results without or with resummation labelled as ``Frag, NLO" and ``Frag, NLO+NLL" are presented in Figs. \ref{gammaz1} and \ref{gammaz31}, which are calculated by using Eqs.(\ref{fragnlo}, \ref{fragnlonll}), respectively. The choices of the factorization and renormalization scales for the ``Frag, NLO" and ``Frag, NLO+NLL" calculations have been given below Eq.(\ref{fragnlo}) and Eq.(\ref{fragnlonll}) accordingly. Here the NLO fixed-order results labelled as ``FO, NLO" are presented as a comparison. The decay widths for ``Frag, NLO" and ``Frag, NLO+NLL" are presented in Table \ref{tb.wbcFrag}. One may observe that by resuming the next-to-leading logarithms of $m_{Q}^2/m_{_W}^2$, more accurate behavior in large $z$ region can be achieved, and the total decay widths shall be reduced by about $2\%$ for both $B_c$ and $B_c^*$ productions.

In the fixed-order prediction, the large logarithms of $m_{Q}^2/m_{_W}^2$ may appear in specific kinematic region, and the fragmentation approach provides us a way to give a reasonable contribution in this region by resuming all the large logarithms. Thus a combination of those two approaches may be helpful. As an attempt, we combine the NLO results from the fixed-order and fragmentation approaches in the following way,
\begin{eqnarray}
&& d\Gamma_{W^+ \to B^{(*)}_c+X}^{\rm Combined, NLO+NLL} \nonumber \\
&=& d\Gamma_{W^+ \to B^{(*)}_c+X}^{\rm FO,NLO} + \left(d\Gamma_{W^+ \to B^{(*)}_c+X}^{\rm Frag, NLO+NLL}-d\Gamma_{W^+ \to B^{(*)}_c+X}^{\rm Frag, NLO}\right). \nonumber
\end{eqnarray}
Here, the choices of the factorization and renormalization scales for the ``Frag, NLO" and ``Frag, NLO+NLL" calculations have been given below Eq.(\ref{fragnlo}) and Eq.(\ref{fragnlonll}) accordingly, and the renormalization for the ``FO" calculation is set as $\mu_R=2m_b$. The differential decay widths $d\Gamma/dz$ for ``Combined, NLO+NLL" are also presented in Figs. \ref{gammaz1} and \ref{gammaz31}. The total decay widths are
\begin{eqnarray}
\Gamma^{\rm Combined, NLO+NLL}_{W^+ \to B_c+X}  &=& 4.78\,{\rm keV},  \label{com1} \\
\Gamma^{\rm Combined, NLO+NLL}_{W^+ \to B^*_c+X}&=& 3.47\,{\rm keV}.  \label{com2}
\end{eqnarray}
The combined results are the definitive results in the presented paper because the large logarithms of $m_Q^2/m_{_W}^2$ have been resummed and the theoretical uncertainties are reduced compared to the fixed-order results.

\subsection{Uncertainty analysis}

In this subsection, we shall estimate the theoretical uncertainties for these decay widths. The main uncertainty sources for these decay widths include the heavy quark masses ($m_c$ and $m_b$), the renormalization scale and the radial wave function at the origin $\vert R_S(0)\vert$. $\vert R_S(0)\vert^2$ is an overall factor in the calculation, the uncertainty due to it can be figured out easily. So we shall not consider the uncertainty of $\vert R_S(0)\vert^2$ and concentrate our attention on the uncertainties from the heavy quark masses and the renormalization scale. For clarify, when considering the uncertainty caused by one input parameter, the other input parameters are fixed to their center values.

\begin{table}[htb]
\begin{tabular}{l c c c}
\hline
 $m_c$(GeV)   & ~~1.4 ~~& ~~1.5~~ & ~~1.6 ~~  \\
\hline
$B_c$(FO,NLO)    & 4.95  &  4.89 & 4.84 \\
$B_c$(Comb,NLO+NLL)    & 4.84  & 4.78  & 4.74 \\
$B_c^*$(FO,NLO)  & 3.57  &  3.55 & 3.53 \\
$B_c^*$(Comb,NLO+NLL)    & 3.50  & 3.47  & 3.46 \\
\hline
\end{tabular}
\caption{The decay widths (in unit: keV) for $W^+\to B_c (B^*_c)+b+\bar{s}+X$ with a variation of $m_c=1.5\pm 0.1\,{\rm GeV}$.}
\label{tb.wbcmc}
\end{table}

\begin{table}[htb]
\begin{tabular}{l c c c}
\hline
 $m_b$(GeV)   & ~~4.7 ~~& ~~4.9~~ & ~~5.1 ~~  \\
\hline
$B_c$(FO,NLO)    & 5.71  &  4.89 & 4.24 \\
$B_c$(Comb,NLO+NLL) & 5.58  &  4.78 & 4.15 \\
$B_c^*$(FO,NLO)  &  4.14 &  3.55 & 3.09 \\
$B_c^*$(Comb,NLO+NLL) &  4.05 &  3.47 & 3.02 \\
\hline
\end{tabular}
\caption{The decay widths (in unit: keV) for $W^+\to B_c (B^*_c)+b+\bar{s}+X$ with a variation of $m_b=4.9 \pm 0.2\,{\rm GeV}$.}
\label{tb.wbcmb}
\end{table}

We first consider the uncertainties caused by the heavy quark masses. We estimate the uncertainties by varying the heavy quark masses with  $m_c=1.5 \pm 0.1\,{\rm GeV}$ and $m_b=4.9 \pm 0.2\,{\rm GeV}$. The uncertainties caused by $m_c$ and $m_b$ are presented in Tables \ref{tb.wbcmc} and \ref{tb.wbcmb} respectively. For comparison, the fixed-order results and the combined NLO+NLL results are presented explicitly. From the two tables, we can see that the decay widths decrease with the increment of $m_c$ and $m_b$, but they are more sensitive to $m_b$ than $m_c$. Adding these two uncertainties caused by $m_c$ and $m_b$ in quadrature, we obtain the uncertainties caused by the heavy quark masses
\begin{eqnarray}
&& \Gamma^{\rm FO, NLO}_{W^+ \to B_c+X}  = 4.89^{+0.82}_{-0.65}\,{\rm keV},   \nonumber \\
&& \Gamma^{\rm FO, NLO}_{W^+ \to B^*_c+X}  = 3.55^{+0.59}_{-0.46}\,{\rm keV}, 
\end{eqnarray}
and
\begin{eqnarray}
&& \Gamma^{\rm Combined, NLO+NLL}_{W^+ \to B_c+X}  = 4.78^{+0.80}_{-0.63}\,{\rm keV},   \nonumber \\
&& \Gamma^{\rm Combined, NLO+NLL}_{W^+ \to B^*_c+X}= 3.47^{+0.58}_{-0.45}\,{\rm keV}.  
\end{eqnarray}

Then we consider the uncertainties caused by the renormalization scale and the factorization scale. For the fixed-order results under the conventional scale setting, the uncertainties caused by the renormalization scale can be estimated by varying the renormalization scale between the two typical energy scales $2m_b$ and $m_{_W}$. The fixed-order results with $\mu_R=2m_b$ and $\mu_R=m_{_W}$ have been presented in Tables \ref{tb.wbcwidth} and \ref{tb.wbc*width}, i.e., $ \Gamma^{\rm FO, NLO}_{W^+ \to B_c+X}\in [2.72,4.89]\,{\rm keV}$ and $\Gamma^{\rm FO, NLO}_{W^+ \to B_c^*+X} \in [2.15, 3.55]\,{\rm keV}$ with a variation of $\mu_R\in [2m_b,m_{_W}]$. These uncertainties caused by the choice of the renormalization scale in the fixed-order results are big. 

For the combined NLO+NLL results, there are several renormalization and factorization scales involved in the calculation. We can decompose the combined NLO+NLL results into fragmentation contribution $d\Gamma_{W^+ \to B^{(*)}_c+X}^{\rm Frag, NLO+NLL}$ and power-correction contribution $(d\Gamma_{W^+ \to B^{(*)}_c+X}^{\rm FO,NLO}-d\Gamma_{W^+ \to B^{(*)}_c+X}^{\rm Frag, NLO})$. For the fragmentation contribution, there are lower factorization and renormalization scales, and upper factorization and renormalization scales involved in the calculation. The lower scales should be ${\cal O}(m_{Q})$ and the upper scales should be ${\cal O}(m_{_W})$. Since we adopt the approximation used in Ref.\cite{MELE} where $\int_0^1 dz P_{c \to c}(z)=0$, the fragmentation probabilities for $c\to B_c(B_c^*)$ do not change under the DGLAP evolution. Thus, for the total decay widths, we need not to consider the uncertainties caused by the lower and upper factorization scales, and only need to consider the uncertainties caused by the lower and upper renormalization scales. For the power-correction contribution, the renormalization scales in $d\Gamma_{W^+ \to B^{(*)}_c+X}^{\rm FO,NLO}$ and $d\Gamma_{W^+ \to B^{(*)}_c+X}^{\rm Frag, NLO}$ should be the same. We take the renormalization scales in the power-correction contribution as the same as the lower renormalization scale in the fragmentation contribution for simplicity. In order to estimate the uncertainties for the combined NLO+NLL results, we vary the lower and upper renormalization scales by a factor of 2 from their center values. 

The decay widths with lower renormalization scale $\mu_{R0}=2m_b$ and $\mu_{R0}=4m_b$ are presented in Table \ref{tb.lowmuR}, and the decay widths with upper renormalization scale $\mu_{R}=m_{_W}$ and $\mu_{R}=m_{_W}/2$ are presented in Table \ref{tb.upmuR}. More explicitly, we obtain $ \Gamma^{\rm Combined, NLO+NLL}_{W^+ \to B_c+X}\in [3.93,4.78]\,{\rm keV}$ and $\Gamma^{\rm Combined, NLO+NLL}_{W^+ \to B_c^*+X} \in [2.97, 3.47]\,{\rm keV}$ with a variation of the lower renormalization scale $\mu_{R0}\in [2m_b,4m_b]$, and $ \Gamma^{\rm Combined, NLO+NLL}_{W^+ \to B_c+X}\in [4.78,4.81]\,{\rm keV}$ and $\Gamma^{\rm Combined, NLO+NLL}_{W^+ \to B_c^*+X} \in [3.47, 3.49]\,{\rm keV}$ with a variation of upper renormalization scale $\mu_{R}\in [m_{_W}/2,m_{_W}]$.

\begin{table}[htb]
\begin{tabular}{l c c }
\hline
 $\mu_{R0}$   & ~~$2m_b$ ~~& ~~$4m_b$~~   \\
\hline
$B_c$   & 4.78  &  3.93  \\
$B_c^*$ &  3.47 &  2.97  \\
\hline
\end{tabular}
\caption{The ``combined, NLO+NLL" decay widths (in unit: keV) with a variation of the lower renormalization scale $\mu_{R0}$.}
\label{tb.lowmuR}
\end{table}

\begin{table}[htb]
\begin{tabular}{l c c }
\hline
 $\mu_{R}$   & ~~$m_{_W}$ ~~& ~~$m_{_W}/2$~~   \\
\hline
$B_c$   & 4.78  &  4.81 \\
$B_c^*$ &  3.47 &   3.49 \\
\hline
\end{tabular}
\caption{The ``combined, NLO+NLL" decay widths (in unit: keV) with a variation of the upper renormalization scale $\mu_{R}$.}
\label{tb.upmuR}
\end{table}

The results in Tables \ref{tb.lowmuR} and \ref{tb.upmuR} show that the combined NLO+NLL results are more sensitive to the lower renormalization scale than the upper renormalization scale. Comparing the results in Tables \ref{tb.lowmuR} and \ref{tb.upmuR} with those in Tables \ref{tb.wbcwidth} and \ref{tb.wbc*width}, we can see that after resumming the large logarithms of $m_Q^2/m_{_W}^2$, the uncertainties caused by the choice of the renormalization scale are reduced explicitly.

\section{Summary}
\label{secCon}

In the present paper, we have calculated the $W^+$-boson decays, $W^+\to B_c (B^*_c)+b+\bar{s}+X$, up to NLO QCD corrections under the NRQCD framework. Both the fixed-order and fragmentation approaches are adopted for the calculations. The theoretical uncertainties are analyzed through varying the heavy quark masses and the renormalization scale. Our results show the NLO corrections are significant. Under conventional scale-setting approach, the decay widths for $W^+\to B_c (B_c^*)+b+\bar{s}+X$ shall be increased by $69\%$ $(43\%)$ for the case of $\mu_R=2m_b$ after including the NLO corrections. The scale dependence can be suppressed after including the NLO corrections, even though it is still large. By using the PMC, we show that the renormalization scale ambiguity can be eliminated for those two decay processes; thus they provide another successful applications of the PMC.

The differential distributions $d\Gamma/dz$ for the $W^+$-boson decays $W^+\to B_c (B^*_c)+b+\bar{s}+X$ have been given under the fixed-order and the fragmentation approaches, respectively. Our results show that both decay processes are dominated by the fragmentation mechanism, and the differences exist in small $z$ region. By combining the fixed-order prediction with the fragmentation approach to resum the leading and next-to-leading logarithms of $m^2_{Q}/m^2_{_W}$, a more accurate distribution to the fixed-order prediction can be achieved.

The total $W$-boson decay width is $\Gamma_{_W}=2.09{\rm GeV}$~\cite{pdg}. Using the combined results (\ref{com1}, \ref{com2}) from the fixed-order and fragmentation approaches, we obtain the branching fractions for the considered channels, e.g.
\begin{eqnarray}
\Gamma_{W^+\to B_c+b+\bar{s}+X}/\Gamma_{_W} &=& 2.29\times 10^{-6}, \\
\Gamma_{W^+\to B^*_c+b+\bar{s}+X}/\Gamma_{_W} &=& 1.66\times 10^{-6}.
\end{eqnarray}
If the LHC runs with a luminosity of $10^{34}{\rm cm}^{-2}{\rm s}^{-1}$~\cite{WBc1}, the expected $W^+$-boson events is about $3.07\times 10^{10}$ events per operation year. Then, there are about $7.03\times 10^4$ $B_c$ mesons and $5.10\times 10^4$ $B_c^*$ mesons to be produced through the $W^+$-boson decays per operation year. Two $2S$-level excited states $B_c(2^1S_0)$ and $B_c(2^3S_1)$ have recently been observed by CMS and LHCb collaborations~\cite{Bc2S2, Bc2S3}. Using $\vert R_{2S}(0)\vert^2=0.983\,{\rm GeV^3}$~\cite{pot}, there are about $4.21\times 10^4$ $B_c(2^1S_0)$ and $3.05\times 10^4$ $B_c^*(2^3S_1)$ to be produced through the $W^+$-boson decays per operation year. Those excited states can also be important sources for the ground-state $B_c$ meson. Thus by carefully measuring the $W^+$-boson decays, we may have a good chance to study the $B_c$ meson properties.

\hspace{2cm}

\noindent {\bf Acknowledgments:} This work was supported in part by the Natural Science Foundation of China under Grant No.11625520, No.11847222, No.11847301, No.11675239, No.11535002, No.11745006, and No.11821505, by the Fundamental Research Funds for the Central Universities under Grant No.2019CDJDWL0005, by the China Postdoctoral Science Foundation under Grant No.2019M663432, by the Chongqing Special Postdoctoral Science Foundation under Grant No.XmT2019055, and by the graduate research and innovation foundation of Chongqing, China under Grant No.CYB19065.

\end{document}